# MAKING SPACE TO SENSEMAKE: EPISTEMIC DISTANCING IN SMALL GROUP PHYSICS DISCUSSIONS

LUKE D. CONLIN
Graduate School of Education, Stanford University
430 Wallenberg Hall Room, Building 160
450 Serra Mall, Stanford, CA 94305
lconlin@stanford.edu

RACHEL E. SCHERR
Department of Physics, Seattle Pacific University
Otto Miller Hall Room 110, 3307 Third Avenue West, Seattle, WA 98119

**This manuscript is currently under review.**

Making Space to Sensemake: Epistemic Distancing in Small Group Physics Discussions

**Abstract**

Students in inquiry science classrooms face an essential tension between sharing new ideas and critically evaluating those ideas. Both sides of this tension pose affective risks that can discourage further discussion, such as the embarrassment of having an idea rejected. This paper presents a close discourse analysis of three groups of undergraduate physics students in their first discussions of the semester, detailing how they navigate these tensions to create a safe space to make sense of physics together. A central finding is that students and instructors alike rely on a common discursive resource – epistemic distancing – to protect affect while beginning to engage with ideas in productive ways. The groups differ in how soon, how often, and how deeply they engage in figuring out mechanisms together, and these differences can be explained, in part, by differences in how they epistemically distance themselves from their claims. Implications for research include the importance of considering the coupled dynamics of epistemology and affect in collaborative sensemaking discussions. Implications for instruction include novel ways of encouraging classroom discussion.

**Making Space to Sensemake: Epistemic Distancing in Small Group Physics Discussions**

Science is driven by an essential tension between two sorts of processes: generative processes of coming up with ideas and communicating them to others, and critical processes of evaluating those ideas and pruning them (T.S. Kuhn, 1977; Popper, 2005). This tension arises in science classrooms, particularly during collaborative scientific sensemaking discussions in which students come up with ideas to explain physical phenomena, share these ideas with others, and critically evaluate each other's ideas (Ford, 2008). Students in these discussions must continually make repairs of each other's understanding, and so must find ways to manage the affective risk of disagreeing if they are to prevent the discussion from shutting down. On the one hand, too much disagreement could discourage further contributions to the discussion. On the other, too *little* disagreement can mean that students are avoiding conflict at the expense of sensemaking together.

Inquiry-based science classrooms can support students in learning through sensemaking discussions (e.g., Roseberry, Warren, Conant, & Hudicourt-Barnes, 1992; Scherr & Hammer, 2009), including charged discussions in which they manage affective risks (Duschl, 2008; Engle & Conant, 2002). But how do the students manage these affective risks in co-creating a safe space to sensemake? And how can these environments support students in managing affective risk of scientific discussions? This paper provides evidence of one important resource students and teachers alike rely on in making a safe space to sensemake: *epistemic distancing*. Speakers use hedging, quoting, questioning, and joking to epistemically distance themselves from their claims, leaving room to evaluate the ideas rather than the person coming up with them. This can contribute to a safe space where the generation and critique of ideas are welcomed, rather than discouraged.

This paper presents an analysis of the early discussions of three groups of undergraduate students working together in introductory physics tutorials, with the goal of understanding how the groups construct a safe space to sensemake. The tutorials are weekly discussion sessions where students meet in groups of four for 50-minutes of worksheet-guided inquiry, as part of their introductory algebra-based physics course. Tutorials are meant to support students in collaboratively making sense of topics in physics that research has shown are particularly challenging for students (Shaffer & McDermott, 1992).

Two tutorial groups' contrasting approaches to a particular physics problem illustrate the idea of a safe space for collaborative sensemaking. The problem comes from the 9$^{th}$ week of tutorial, in which the students are exploring the physics of how a roller coaster cart can make it all the way around a vertical loop in the track. Students are asked to draw a diagram showing all the forces acting on a roller coaster cart when it is upside-down at the top of the loop (Figure 1, Point B). The correct answer is that there are two forces: the gravitational force exerted on the cart by the Earth, and the contact force exerted on the cart by the track. When the cart is at point B, both of these forces point vertically downward.

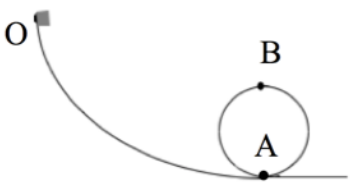

Figure 1—Diagram of roller-coaster track in tutorial problem. Students are asked, "A cart is released at point O, goes down a hill, and goes around a vertical loop in the track. What forces act on the cart when it is at point B?"

The students in the Gold group contribute their ideas about which forces act on the cart. They all agree that gravity is pulling down on the cart. They also agree that the track exerts a force on the cart, but they disagree on its direction:

> CARMELLE: I mean you have the force of the track pressing it down,
> BREE: The force of the track pushing down.
> DEIRDRE: But wouldn't it-
> AMANDA: Going down.
> DEIRDRE: Would it be going up or would it be going like, *(draws an arrow that points up at an angle)* like that?
> BREE: What?
> DEIRDRE: The force of the track.
> BREE: Nah, it's going down.
> CARMELLE: 'Cause it's pressing down on it, it's at- it's at the top part it that top part of is is what's pushing down *(gesture: one hand representing track on top of other representing car, and pushing it down)*
> BREE: *(overlapping with Carmelle)* Pushing down, cause that's what's holding it in *(gesture: pointing index fingers in towards body)* from like being shot like way out *(gesture: pointing away from body with index finger, shooting hand away)*

Carmelle, Bree, and Amanda[1] state that the force on the cart by the track should point downward, but Deirdre contributes the conflicting idea that it should be pointed up, or at an angle. In disagreeing, Deirdre softens her claim by phrasing it as a question, an epistemic distancing move. This opens space for both Carmelle and Bree to reject Deirdre's idea as they provide justification for why the force should be downwards. Carmelle's explanation is in terms of the configuration of the car and the track: The track is on top so it must be "pressing down" on the car. Bree's explanation is in terms of competing influences: If the track is "holding [the cart] in" from "being shot like way out," then the track must be pushing the cart inwards (toward the center of the loop). Both explanations are based on the students' sense of how the track interacts with the cart. The members of the group are willing to contribute their ideas about the physical mechanisms at work and critically evaluate each other's contributions. In other words, they have made a safe space for collaborative scientific sensemaking, in alignment with the goals of the tutorial curriculum.

---

[1] All names are psuedonyms.

The Bronze group, in contrast, does not explicitly share their ideas about which forces are acting on the cart. They have their lecture notes out and are trying to recreate the diagram the lecturer had drawn on the board. Their notes are not consistent with each other, which leads to disagreement:

> CENA: *(flips through lecture notes, then points at Britte's worksheet)* Why does he say minus N? I remember him writing that down, but why is it minus N.
> BRITTE: *(opens lecture notes and reads, then points to Devin's worksheet)* Are you sure that's just like N and MG? Because like, I have like written like N – MG and stuff like he talked about
> DEVIN: Right, that's for- *(trails off)*
> *(everyone reads silently for 30 seconds)*
> BRITTE: *(reading from lecture notes)* "N minus MG equals force directed into the circle."...I guess if you're at point A, isn't that the lower one?
> DEVIN: Yeah I have it written down MG – N = mv2/r at the top.
> BRITTE: At top, so that's right at B
> DEVIN: Yeah
> BRITTE: So that's when you're at B...You're here *(points to top of loop)*
> DEVIN: Yah.
> BRITTE: *(pointing to DEVIN'S notes)* That equation.

The students in the Bronze group refer to the equations in their lecture notes to determine the forces on the cart at the top of the vertical loop. Like the students in the Gold group, they correctly include the force of gravity (which they refer to as "mg" because its magnitude is the product of m, the mass of the object, and g, the gravitational acceleration constant near the surface of the earth) and the force of the track (which they refer to as "N" for "normal force," which is a contact force perpendicular to the surface of contact). Cena asks why the N has a minus sign. Britte has "N – mg" in her notes, while Devin has "mg – N," which seems to imply disagreement about the direction of the normal force. Britte reads her notes more closely to notice that the "N – mg" refers to the point at the bottom of the track (Point A), not at the top (Point B). Devin says, "I have 'mg – N = $mv^2/r$' at the top," which convinces Britte the equation should be "mg – N" at the top of the loop. Thus, the Bronze group does not attempt to resolve their disagreement by making sense of the physical mechanism. Instead, they attempt to make sure their diagram matches what they copied down in their lecture notes. When an instructor comes by later, he is surprised to find that they have not only all drawn the force on the cart by the track in the wrong direction, but also drawn the gravitational force pointing up.

The two groups take very different approaches to the same problem. In the Gold group, students contribute their ideas about the direction of the forces acting on the cart, and justify these ideas based on their sense of the mechanism of interaction between the cart and the track. They also evaluate each other's ideas, risking embarrassment in disagreeing with each other. In the Bronze group, students do not share their ideas about what is going on between the cart and the track. While they do disagree, it is only with respect to figuring out which diagram was drawn where in their notes. In other words, the Gold group establishes a safe space to collaboratively make sense of the interaction between the cart and the track, while the Bronze group does not. The Bronze group's approach is out of alignment with the goals of the curriculum, and has negative consequences for building a shared conceptual understanding of the direction of the forces acting on the cart at the top of the loop.

How do these two tutorial groups come to frame their activity so differently? What leads one group to make sense of the physical scenario, while the other refers only to their lecture notes? Part of the answer can be found within these data examples. In the Gold group, Deirdre softens the blow of her disagreement about the direction of the normal force by phrasing it as a question, which leaves space for Bree and Carmelle to justify their rejection of the idea in terms of their sense of physical mechanism. Thus, Deirdre's epistemic distancing contributes to the Gold group's safe space for sensemaking.

This is not enough to explain the Bronze group's different approach in this instance, however. Britte also phrases her disagreement as a question ("Are you sure that's just like N and MG?"), but this does not lead them to make sense of the physical scenario. The question remains as to why the Bronze group appeals to their lecture notes to settle disagreements, while the Gold group relies on their sense of mechanism. To answer this question requires an examination of each groups' broader history of interactions.

Prior research indicates that the groups' different approaches are established very early on in the semester; the tone is set within their first few discussions (Conlin, 2012). This paper presents a close analysis of the discourse of three tutorial groups' early discussions, to understand how the groups first manage to construct a safe space for sensemaking together. A central finding is that students' use of *epistemic distancing* – hedging, joking, and other discourse moves to soften one's stance in conversation – plays a critical role in each groups' initial construction of a safe space to discuss their ideas about mechanism. Differences in the groups' use of epistemic distancing also help explain variability across the groups in how soon, how often, and how deeply they make sense of mechanisms together.

The next section reviews research on stancetaking in conversation to characterize epistemic distancing and to describe its potential for mitigating conflict within scientific sensemaking discussions.

## Managing Conflict through Epistemic Distancing

Managing affective risk is essential to authentic disciplinary engagement. Critique and skepticism are necessary for building reliable explanations in science (Ford, 2008; Osborne, 2010), but when scientists' ideas are rejected it can do damage to their reputation within the scientific community.[2] In active engagement classrooms where students must resolve conflicts amongst competing ideas face-to-face, these affective risks become even more immediate. Disagreements can cause frustration, embarrassment, and loss of face (Brown & Levinson, 1987; Goffman, 1955, 1956). Students experiencing such repercussions in collaborative group work may become reluctant to contribute more ideas. If groups are to collaborate in scientific argumentation, they must find a way to manage these affective tensions, in addition to the conceptual and epistemological ones (Barron, 2003; Berland & Hammer, 2012; Engle & Conant, 2002).

Little is understood about how instructors and students in science classrooms manage to create a safe space to share and critique ideas, in light of challenges that arise during face-to-face interactions. In what follows, we draw from research on discourse in interaction (Johnstone, 2008; Schiffrin, Tannen, & Hamilton, 2001) to characterize one resource – epistemic distancing

---

[2] This was the case for Dan Shechtman, recipient of the 2011 Nobel Prize in chemistry for the discovery of quasicrystals. When his idea was originally rejected, Shechtman's career was all but ruined. He was expelled by his research group and ridiculed by leading chemists such as Linus Pauling, who quipped: "There is no such thing as quasicrystals, only quasi-scientists."

– by which students and instructors alike may manage affect in sensemaking discussions. By epistemically distancing themselves from their ideas, students can protect themselves from the affective damage that comes with having their ideas critiqued.

**Stance-taking in Conversation**

Students engaged in scientific sensemaking discussions continually make claims, display attitudes, and express evaluations, all of which discourse analysts have described broadly as taking *stances* in conversation (Kärkkäinen, 2006; Kirkham, West, & Street, 2011). Participants can take stances toward the conceptual substance of what they are discussing, i.e., that the force is directed down. They can also take stances toward the source and reliability of the knowledge being expressed, i.e., what discourse analysts have described as *epistemic stance* (Biber, 1989, 2006; Kiesling, 2009). Often, these co-occur. For instance, when Deirdre asks about the force on the cart from the track, "Wouldn't it- Would it be going up…?" she is simultaneously conveying an idea about the direction of the force while conveying uncertainty in this idea by phrasing it as a question.

Speakers can upgrade or downgrade their epistemic stance through various discourse moves, for instance by *deferring* (e.g., "research has proven…") or by *hedging* (e.g., "I guess…") (Clift, 2006; Kärkkäinen, 2007). Any discourse move that either strengthens or weakens a speaker's stance are described as shifts in a speaker's *footing* (Goffman, 1979; Clift, 2006). Footing shifts can be accomplished through explicit hedging using phrases such as "I think" (Kärkkäinen, 2003; Holmes, 1990), but can be conveyed through paralinguistic channels as well. These include the use of a fall-rise intonation to express uncertainty (Ward & Hirschberg, 1985), adoption of sing-song prosody to convey irony (Clift, 1999), or the shift of body posture to broadcast resistance to an idea (Goodwin, 2007a, 2007b).

Discourse analysts have highlighted how people index their stance in conversation to manage conflicts (Bonito & Sanders, 2002; Heisterkamp, 2006; Jacobs, 2002; Kärkkäinen, 2003, 2006; Sharma, 2011). By using the phrase "I think" to soften their stance, speakers can avoid the threat to face that comes with bringing up a controversial topic (Kärkkäinen, 2006). Bonito and Sanders (2002) found that by deferring to each other when disagreements arose, students engaged in a collaborative writing task used footing shifts in ways that allowed them to express contrary positions without escalating the conflict.

**Epistemic Distancing**

In managing face-to-face conflict, footing shifts that serve to soften one's stance play such a critical role that it is worth distinguishing them with their own term – *epistemic distancing*. Epistemic distancing refers to any discourse move by which a participant in conversation downgrades their epistemic stance, for example by hedging, quoting, or joking. Epistemic distancing is "epistemic" in that it concerns the speaker's commitment to the truth of what they are saying. It is "distancing" in that it creates distance between the speaker and what they are saying (Goffman, 1979). This distance protects the speaker's affect in the event that the idea gets evaluated negatively, thereby reducing the ego threat that can discourage further contributions.

While research has not yet explored how students can index their stance to mitigate conflict in science classrooms, the findings from discourse analysis suggest that epistemic distancing can serve as a resource by which students simultaneously manage the epistemological and affective dynamics of conflict resolution. It is by managing these dynamics that the students in tutorial can build and maintain a safe space for collaboratively making sense of physics. To do so, students may need to distance themselves not only from their ideas about physics, but also their ideas about how to go about learning physics together.

For a brief illustrative example of the latter, consider how one student responds to the very first tutorial question: "What do you think are the benefits of discussing your mistakes in physics? Discuss your answers." After an uncomfortable silence, the group decides to read their answers aloud. Bree goes first, reading what she wrote in an ironic, performative manner, with exaggerated pitch variations, gestures, and facial expressions (Figure 2).

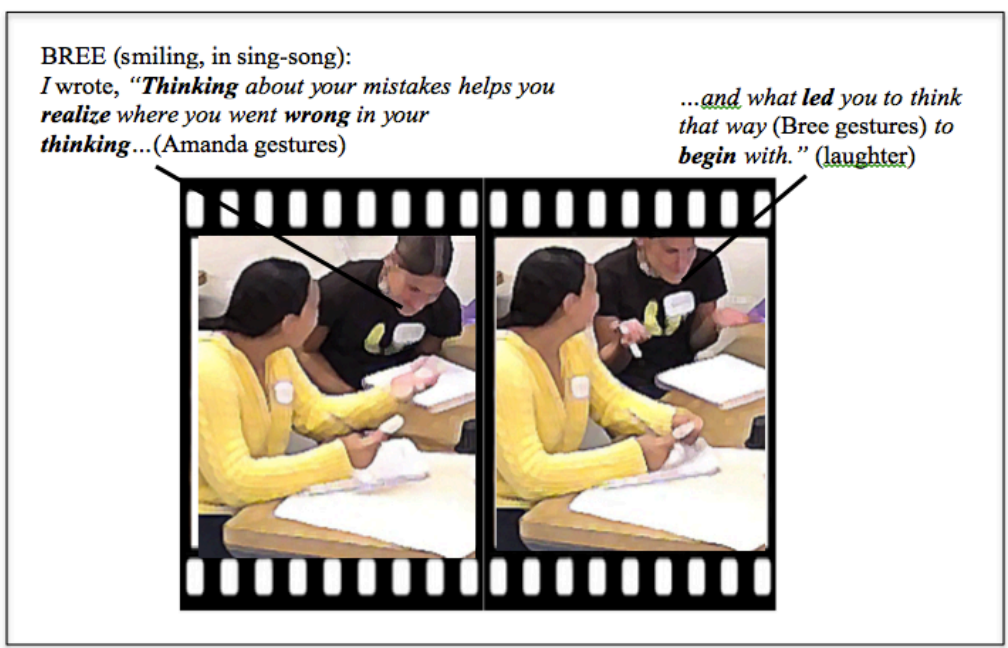

Figure 2—Bree reads her response to the first tutorial question. She softens the stance implied in her written words via an ironic shift in footing, using sing-song prosody and performative gestures (Clift, 2006).

Bree is epistemically distancing herself from what she says in two ways. She adds one layer of distance by reading what she wrote, instead of saying what she thinks. She adds a second layer of distance by reading her response in a sing-song, performative manner. This move constitutes an ironic shift of footing (Clift, 1999) that further downgrades the stance taken in her written response, signaling that she does not take her response too seriously. In the words of Goffman, (1974, p. 512):

> When a speaker employs conventional brackets to warn us that what [s]he is saying is meant to be taken in jest, or as mere repeating of words by someone else, then it is clear that [s]he means to stand in a relation of reduced personal responsibility for what [s]he is saying. [S]he splits [her]self off from the content of the words by expressing that their speaker is not [s]he [her]self or not [s]he [her]self in a serious way.

By expressing her view, but not taking it too seriously, Bree downgrades her epistemic stance in a way that softens any impending conflict that might arise from someone expressing a different view. This in turn can make others more comfortable with sharing their perspective. Thus, through the use of epistemic distancing students can make space for multiple perspectives as they position themselves with respect to contrasting claims. As detailed below, Bree's epistemic distancing helps her group shift away from reading what they wrote and toward saying what they think, an important step on the way to collaborative sensemaking.

This is not to say that more epistemic distancing is necessarily better for collaborative sensemaking discussions. Students who distance themselves too much from their ideas about physics or how to learn physics can end up avoiding conflict completely, or else come across as so dismissive of the activity that they discourage further contributions from the group. The next section presents a close discourse analysis of three tutorial groups' first discussions of the semester, noting when epistemic distancing is happening (or not) and tracking its effect on the dynamics of collaborative sensemaking.

## Data & Analysis

**Instructional setting**
The *Tutorials in Physics Sensemaking* are worksheet-guided discussion sections designed to support inquiry into various topics as part of an introductory algebra-based physics course. At the University of Maryland, where these tutorials were developed and where this study was conducted, students in the algebra-based physics course are primarily life science majors in their junior year. Tutorial groups meet once a week for a 50-minute session of collaborative work. Six teams of four students each are supported by one or two Teaching Assistants (TAs). The worksheets are not collected or graded. The students often do not know each other when they sit down together their on their first day. They may sit wherever they like, but they generally stay with the same groups throughout the semester.

**Data collection**
The video data comes from a large corpus (~2,000 hours) of videotaped tutorial sessions at the University of Maryland, recorded as part of a larger study of students' reasoning during tutorials (see Scherr, 2009). During each tutorial session, the activity at two of the tables was recorded by a pair of small stationary cameras, placed on the periphery of the room. The two tables were equipped with embedded microphones. Seating was not assigned, but since the students tended to keep the same seating arrangements we were able to follow intact groups throughout the semester. Three groups were selected for comparison across a range in levels of engagement with the tutorials, referred to in this paper as the Gold, Silver, and Bronze groups. The Gold and Silver groups were in the same year and same section, and so were in same room during the time of recording. The Bronze group was recorded two years later.

**Episode selection**
To investigate how student groups initially engage in collaborative scientific sensemaking, the analysis focuses on the groups' first few discussions of the semester. First, we used group-level shifts in body positioning to identify the very first time each group orients to a discussion space (McDermott, Gospodinoff, & Aron, 1978; Scherr & Hammer, 2009). For all three groups this happens in response to the very first tutorial question of the semester. Next, we located the first discussion of each group that contains substantial evidence of collaborative scientific sensemaking, which happens in responses to different questions for the different groups. We examine how each group gets into their first discussion in Part I below, and into their first collaborative sensemaking discussion in Part II.

**Data Analysis Part I – 1st discussions**

All three groups get into their first discussion of the semester in response to the instructions of the first tutorial question, which asks them how thinking about their mistakes may help them learn physics (Figure 3). Groups vary in how they take up this discussion. Epistemic distancing helps explain this variability.

> Since reflecting on the purpose of an activity can help you get more out of it, let's start with this:
>
> A. *(Answer individually)* What do you see as potential benefits of explicitly thinking and talking about the mistakes you make while working through these activities? If you think dwelling on your mistakes won't be particularly helpful, explain why not.
>
> B. Discuss your answers with your group. If anyone gave part of an answer significantly different from yours, write a one-sentence-summary of that opinion here.

Figure 3—Part A of Question 1 in Tutorial 1 asks students to reflect on the potential benefits of thinking and talking about mistakes they make. Part B asks them to discuss their responses with their group.

**The Gold Group's 1st discussion—"I guess we should…'discussss our answersss'"**

After the TA's introduction to the tutorials, the Gold group starts the tutorial silently focused on their worksheets. After a few minutes, the group suddenly transitions to discussing their responses to the first tutorial question (Figure 4).

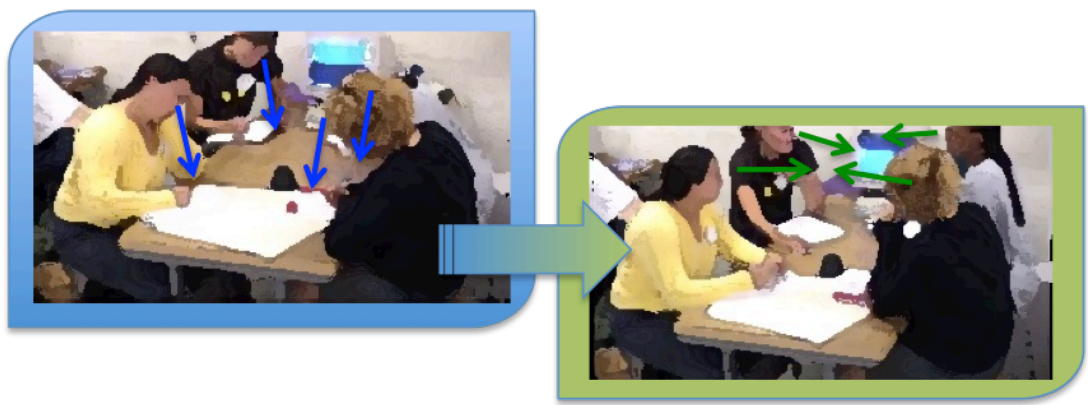

Figure 4 —The Gold Group's gaze shifts during their first transition from completing the worksheet to having a discussion. Clockwise from front left, the members of the Gold group are: Amanda, Bree, Carmelle, and Deirdre.

Behaviorally, each student orients to the group space one at a time over a span of about thirty seconds. Deirdre transitions first. As she finishes Part A she sits back, lifts her hands away from her tutorial worksheet, and looks up. This constitutes an example of what Scherr and Hammer (2009) call a *bid* for a change in activity. After the last student orients to the group space Deirdre says "I guess we should… what is it we have to do?" and the discussion begins:

>           DEIRDRE:   I guess we should...what did we have to do?
>              BREE:   *(in a mocking tone)* "Discusssss our answersss..."

>      AMANDA: I'm sure we all wrote the same thing *(laughs)*
>     DEIRDRE: We could just **read** it to each other, I dunno, to see...
>        BREE: Well...
>      AMANDA: What'd you write, Bree?
>        BREE: *(smiling, in a mocking tone)* I wrote, "**Thinking** about your mistakes helps you realize where you went **wrong** in your **thinking** and what **led** you to think that way *(Bree gestures with open palms)* to **begin** with." *(laughter)*
>      AMANDA: *(laughing)* I wrote exactly the same thing.

Deirdre starts off with a suggestion that they follow the instructions of the tutorial. She epistemically distances herself from the suggestion using the phrase "I guess," conveying uncertainty in her commitment to the suggestion (Kärkkäinen, 2007). Bree answers Deirdre's question about what they were supposed to do by reading the instructions from the worksheet, but with elaborated enunciation: *"Discussss our answerssss."* Bree's mocking tone signals an ironic shift of footing (Clift, 1999) that allows her to express the instructions while epistemically distancing herself from the commitment to following them. Amanda offers further resistance to discussion by saying, "I'm sure we all wrote the same thing," which would apparently obviate the instructions of the tutorial. Deirdre says, "We could just **read** it to each other I dunno, to see…" Reading out loud constitutes a shift in footing that allows the students to minimally follow the tutorial instructions while also distancing themselves from their responses. In this way, the Gold group establishes a precedent of taking the tutorial seriously, but not *too* seriously.

Bree reads her response first, but performs her reading of it with a smirk, and with exaggerated pronunciation, prosody, and gesture. She is apparently poking fun at what she wrote by playing as if she is "teaching" it to the others, an ironic shift in footing that allows Bree to express her idea about learning from mistakes, while at the same time epistemically distancing herself from what she has written. While Bree is reading, Amanda laughs and plays along, expressing agreement with two open palms (Fig. 2) before replying, "I wrote exactly the same thing" in a similar register and laughing. After Bree's turn, Carmelle starts to read her response:

>    CARMELLE: I just put that it um,
>        BREE: ...silly.
>    CARMELLE: Oh, you still goin' I'm sorry
>        BREE: Oh nonono I'm done
>    CARMELLE: I was just gonna say it comforts others in knowing that they too may have made the same mistakes, so you don't feel like you're alone, *(Bree nods)* and um, I also said it kind of fosters better reasoning because *(looks up)* if you can reason through you mistakes then you can-
>  TA ROSSLYN: *(off camera)* Real quick, guys, I 'm sorry to inter- I need to explain to you about how to do the experiment for this one...

Carmelle starts reading her response with a bit of epistemic distancing, prefacing with "just" in "I just put..." and "I was just gonna say…" She reads her response in earnest, without a mocking tone. When she introduces the idea that discussing mistakes can be comforting in that "you don't feel like you're alone," Bree nods in agreement. By the end of the turn (before the TA interrupts with instructions for the class), Carmelle is no longer reading from her worksheet but is looking up and is saying what she thinks.

In summary, the members of the Gold group use epistemic distancing in ways that help them take the tutorial question seriously, without taking it *too* seriously. Deirdre makes a bid to engage with the tutorial, but softens her bid first by hedging ("I guess we should") and then by turning it into a question ("What is it we have to do?"). Bree answers Deirdre's question ("Discussss our answersss") with exaggerated pronunciation, distancing herself from the content of her suggestion to discuss their answers. Deirdre distances them further from the task by suggesting they read their responses. Bree reads her response ironically, allowing her to express her idea while protecting herself with epistemic distancing. When Carmelle takes her turn, she does not need as much epistemic distancing to say what she thinks. Overall, the Gold group uses epistemic distancing to move towards saying what they think, an important step towards having collaborative sensemaking discussions about physics, as will be discussed in Part II.

**The Silver group's 1st discussion—"Whatever…next!"**

Like the Gold group, the Silver group starts the tutorial by reading the worksheet, then transitions together into behaviorally orienting to a discussion space (Figure 5). The entire transition takes thirty seconds.

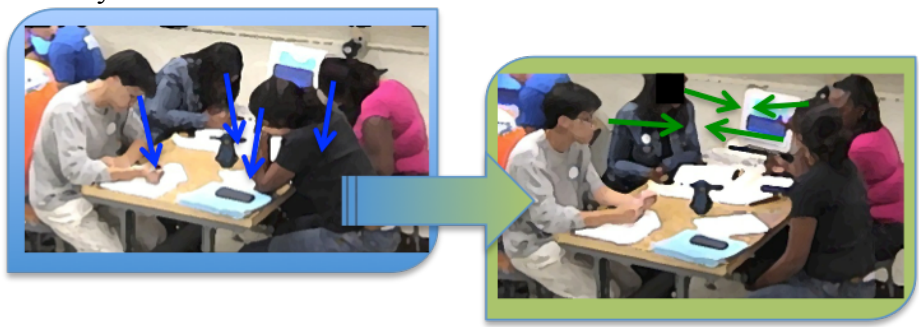

Figure 5—The Silver group's gaze shifts during their first transition from completing the worksheet to having a discussion. Clockwise from front left, the members of the Silver group are: Alan, Brandi, Chrissie, and Daria.

Their discussion of the first question is much more brief than the Gold Group's. Daria is the first to speak, but instead of reading her response, she speaks in generalities:

```
CHRISSIE:  (laughs)
   DARIA:  So...okay...we talked about how you can learn from your mistakes
           pretty much yeah
    ALAN:  Yeah I think everyone said "learning from your mistakes," right?
   DARIA:  Yeah
  BRANDI:  Right
CHRISSIE:  (laughs)
   DARIA:  pretty much okay
    ALAN:  Whatever...next!
```

Daria epistemically distances herself from her contribution in multiple ways. Instead of discussing her idea specifically, she keeps it general. Her use of the pronoun "we" instead of "I" constitutes a shift of footing that introduces distance between her and her idea. Her contribution, "you can learn from your mistakes," does not offer much beyond the question prompt, and she punctuates it with a hedge, "pretty much yeah." Alan endorses the generality of her contribution, also attributing it to the whole group ("I think everyone said [that], right?") rather than sharing his own thoughts. Chrissie laughs and Alan closes the brief discussion with a dismissal, "Whatever…next!"

The Silver group engages with the substance of the question much more superficially than does the Gold group. Nobody in the Silver group actually reads their response, or takes personal responsibility for a contribution to the discussion. In this case, the Silver group distanced themselves too much to engage productively in the discussion, as exemplified by Alan's "Whatever…next!" At this point, the group could be in danger of aligning against the grain of the tutorial's goals. They will continue on in this direction until a TA's intervention to be discussed in Part II.

**The Bronze group's 1st discussion—"It's been proven that you learn from your mistakes"**

Like the Gold group and the Silver group, the Bronze group orients to the group space after an extended period of focusing on their individual worksheets (Figure 6).

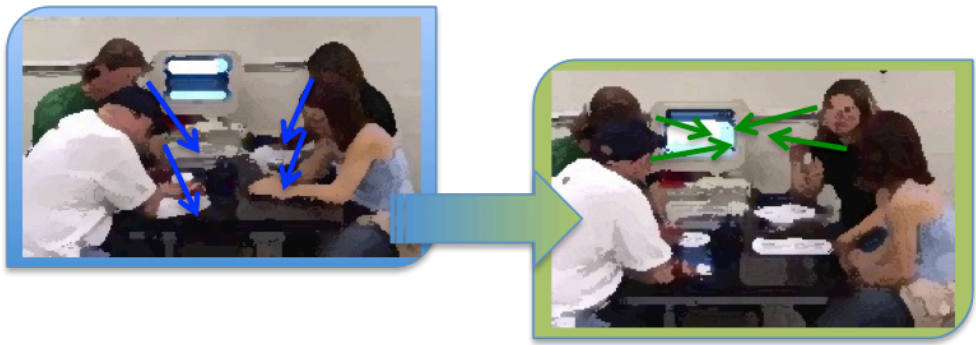

Figure 6—The Bronze group's gaze shifts during their first transition from completing the worksheet to having a discussion. Clockwise from front left, the members of the Bronze group are: Alan, Brad, Cathy, and Devin.

Adam is the first to transition in his behaviors when he puts his pen down and looks up at the computer screen. He apparently finishes responding to the tutorial question about a minute before anyone else. Towards the end of this minute, Brad makes a disparaging comment on the tutorial question right before Cathy looks up and starts the discussion.

```
   BRAD:  PShshss this is very...condescending
  CATHY:  What were...your reasons?
  DEVIN:  So just allows you to better understand...the way you thought about
          it=
  CATHY:  I said...if you
  DEVIN:  =versus the correct way, so you can sorta be able to assess the
          situation better next time.
  CATHY:  Yeah, if you- can catch your mistakes you might notice like a
          pattern of what you- like, what topic you're not understanding
   ADAM:  It's been proven that you learn from your mistakes.
   BRAD:  M'yah.
```

Despite Brad's disparaging comment, the group enters the discussion by tacitly agreeing to follow the instructions. While Cathy and Devin each share their ideas about how discussing mistakes can help them learn, Adam states matter-of-factly: "It's been proven that you learn from your mistakes." In his statement, Adam shifts to the passive construction "it's been proven…", a move that strengthens his stance by deferring to authoritative findings (Clift, 2006). This statement leaves very little room for disagreement, both because of the lack of hedging and because it is a general statement with which everyone most likely would agree (as was the case for the Silver Group). This allows Brad, who had already expressed displeasure with the activity, to simply agree with a "M'yah" without sharing his own ideas. This is a case where lack of epistemic distancing seems to have shut down the conversation.

### Summary of Part I – Making Space for Discussion

Part I examined each tutorial group's very first discussion, bracketed by their behavior orientation to the group space. There is variability in how deeply the groups engage in discussing their ideas about the first tutorial question, which asks what they think the benefits are of discussing their mistakes. Differences in the groups' use of epistemic distancing help explain this variability. The Bronze group's discussion was cut short by a statement with very little epistemic distancing. In contrast, the Silver group's discussion was preempted by too much distancing ("Whatever…next!"). The Gold Group were able to use just enough epistemic distancing to allow them to make fun of the tutorial and even their own answers while easing into saying what they think.

These contrasting cases reveal that epistemic distancing is not unilaterally beneficial to opening up space for discussion; there can be "too much" distancing. Determining a productive amount of epistemic distancing is an empirical matter, to be decided based on whether a group makes progress towards sharing and evaluating each others' ideas, paired with a close analysis of the role that stance-taking plays in that dynamic. The analysis of Part I illustrates the critical role of epistemic distancing in these groups' construction of a discussion space, an important step towards collaborative scientific sensemaking, which will be pursued the focus of the analysis in Part II.

## Data Analysis Part II – 1$^{st}$ collaborative sensemaking discussions

In this second analysis, we identify the dynamics by which each group first succeeds in making space to collaboratively sensemake. The analysis focuses on the first discussion for each group that includes evidence of students contributing and evaluating ideas about physical mechanisms (Russ, Scherr, Hammer, & Mikeska, 2009). For each group, this happens at different times, in response to different tutorial questions (Table 1). In each case, however, the students' and instructors' use of epistemic distancing plays a critical role.

| Tutorial Group | # of discussions until evidence of sensemaking | Elapsed time until evidence of sensemaking | Tutorial question where sensemaking occurred |
|---|---|---|---|
| Gold group | 3 | 13:15 | Tutorial 1, Question II.B.1 |
| Silver group | 3 | 12:30 | Tutorial 1, Question II.A.4 |
| Bronze group | 4 | 52:46 | Tutorial 2, Question 1.A.I |

Table 1. Variability in how soon each group enters a collaborative scientific sensemaking discussion, as measured in number of discussions, elapsed time, and tutorial question.

**The Gold Group's 1st collaborative sensemaking discussion**

The Gold group started making sense of mechanisms soon into the first tutorial. Their third discussion contained evidence of collaborative scientific sensemaking, in response to the third tutorial question. The second question had asked students to stand 0.5 meters away from a motion detector and walk slowly away as it makes a plot of their distance from the detector as a function of time. The third question asks them predict what the graph would look like if they started at one meter away and walk away faster, individually recording their predictions by drawing a dotted line on their graph then discussing to come to a consensus graph. Carmelle expresses confusion over the "dotted line thing", and they discuss:

```
CARMELLE:  Darn it! Why am I not doing this dotted line thing?
    BREE:  So it'd just be like a steeper slope (gestures straight line with pen)
  AMANDA:  Right, okay.
 DEIRDRE:  Steeper slope, that's what- okay.
  AMANDA:  And not starting at the origin
 DEIRDRE:  Yeah a little bit higher
  AMANDA:  Yeah
 DEIRDRE:  (reading) and then, same thing (starts to write)
CARMELLE:  But you know what...(they all look at her) Okay. Okay. Okay right
           cause the steeper slope would represent=
  AMANDA:  (over Carmelle) Going faster
 DEIRDRE:  (over Carmelle) A shorter
CARMELLE:  =a farther distance in shorter time (Amanda and Bree say "shorter
           time" in unison with her) Okay
  AMANDA:  Right.
CARMELLE:  Okay. (nods)
```

In collaboratively predicting what the graph will look like, the Gold group contributes ideas to explain why it will look like that, and critically evaluates those ideas. Bree suggests the slope of the graph will be steeper; Amanda and Deirdre agree. Carmelle seems poised to disagree ("But you know what…") but then immediately softens her stance and finds agreement with the idea. In resolving her potential agreement, she offers a conceptual justification for the idea: "the steeper slope would represent a father distance in a shorter time." Amanda confirms with a "Right" and the group agrees on their graph. From this point on, the Gold group continues to collaboratively make sense of mechanisms regularly throughout the semester.

**The Silver group's 1st collaborative sensemaking discussion**

The Silver group's initially dismissive approach continued for the group's subsequent discussions, until later in Tutorial 1 when a TA overhears them dismissing what he thinks is a good question. The TA uses this as an opportunity to get the Silver group sensemaking. In his prompting, he incrementally adds epistemic distancing into his questioning pattern until the students start coming up with ideas. Then, he uses this sensemaking discussion to make repairs to the Silver group's understanding of what it is they are supposed to be doing in tutorial. After the TA leaves, they continue to collaboratively sensemake without him. I will describe the TA's intervention as well as the Silver group's discussion after the TA leaves.

The Silver group is working on second section of the first tutorial, which asks a student to walk slowly and steadily away from a motion detector, making a graph of the student's distance from the detector as a function of time. The students in the Silver group have all predicted a straight line with a positive slope, depicting the walker's distance from the detector steadily increasing with time.

Alan is the walker for this experiment. He walks slowly and steadily backward, holding a book out in front of him as a target for the motion detector. As he is returning to the table after making the graph, he notices two "jumps" in the graph that deviate from the straight line:

    ALAN: Wh- what are those two jumps?
    DARIA: *(laughing)* Heh- I don't know.
    ALAN: Whatever. *(Sits down)*
    CHRISSIE: Okay, *(reading out loud and trailing off)* "Sketch the result"…
    DARIA: *(trailing off)* You wanna try it again?
    CHRISSIE: *(reading out loud and trailing off)* "Sketch the result"…

A TA (Joey) overhears Alan's question and dismissal and joins their discussion, saying, "So wait a second, that's a- that's a good question. What are those two jumps?" (Figure 7)

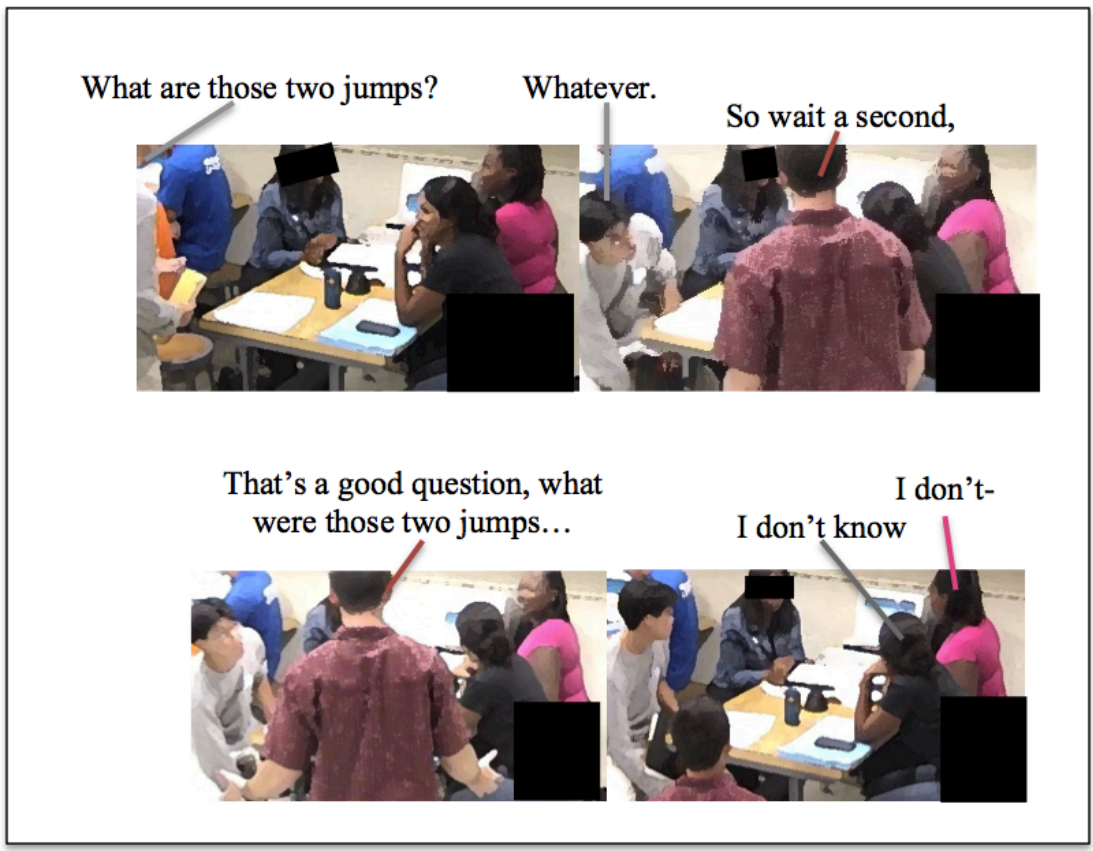

Figure 7—TA Joey overhears the group dismissing a good question and joins in to help the Silver group make sense of the graphs.

When nobody responds, TA Joey kneels down and asks the question again, but with some epistemic distancing (Figure 8).

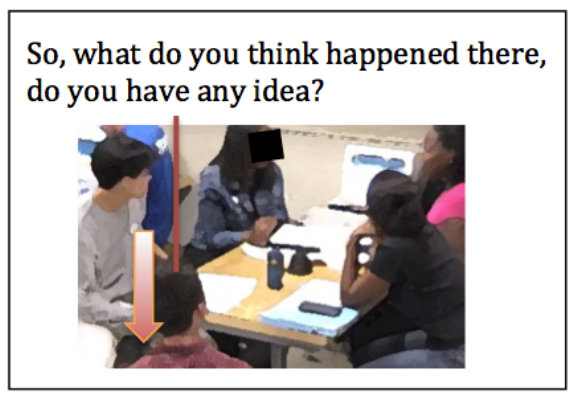

Figure 8—TA Joey introduces more epistemic distance to the question, when he kneels down and asks about what they *think* happened there.

TA Joey introduces epistemic distance in his question using both linguistic and paralinguistic channels. Instead of asking "What happened there," this time he asks "What do you *think* happened there, *any idea*?" (emphasis added). By asking what they *think* TA Joey introduces a hedge that lowers the stakes for contributing ideas without being certain, as does his move to ask if they have *any idea*. His rephrasing of the question invites students to offer ideas even if they do not *know* what happened there. TA Joey also introduces a rising intonation to his question, conveying more uncertainty than before. Finally, TA Joey kneels down as he asks the question, bringing him from an authoritative "hovering" stance to a position in which he is below the students, looking up at them. All of these subtle moves contribute to a safe space in which the Silver group is willing to share their ideas to explain the jumps in the graph:

> TA JOEY: What do you think happened there, do you have any idea?
> ALAN: Ahhh...
> TA JOEY: Because your- It looks like everyone's prediction was a straight line
> DARIA: Right
> TA JOEY: Right? And then, it's mostly a straight line *(gestures out the shape of straight line with two hands)*, but, not exactly. So what's-
> DARIA: Something wrong must've happened.
> ALAN: I dunno. Maybe, this was weird?
> DARIA: Hehehe
> TA JOEY: Maybe it was weird.
> ALAN: Yeah, or
> TA JOEY: What do you mean by 'mayb-' "Weird" could mean a lot of things.
> DARIA: Maybe it's just getting started up or something.
> CHRISSIE: HaHAha!
> TA JOEY: It was getting started up, so like if we did it again *(rolling hand motion)*, like now it's warmed up almost
> DARIA: Mayyybe
> ALAN: Maybe
> CHRISSIE: I think we should do a second trial, to see
> BRANDI: M'yah, maybe he wasn't walkin' that steady

```
   CHRISSIE:  Right. At a steady pace,
      DARIA:  Oh that could be
   TA JOEY:  So this is the sort of thing we want you to investigate. You know,
             like this MOSTly fits with your prediction, but there's some
             discrepencies, and what are they, can you explain why, or maybe,
             like you were sayin' "We wanna try it again." Well, inVEStigate those
             things, don't just say, "Oh, it's exactly what we thought." Because it's
             NOT, quite.
     BRANDI:  Right.
       ALAN:  Okay.
      DARIA:  Okay.
```

Here TA Joey is engaging the Silver group in a sensemaking discussion about what might be causing the jumps in the graph. The students offer competing suggestions, such as an unsteady walking pace and an inadvertent movement of the book they were using as target for the motion detector. The Silver group is using considerable epistemic distancing as they introduce their ideas, with hedges such as "maybe" and "I think." They are also laughing as the ideas are introduced, hinting that they may be half-joking. Alan suggests, "Maybe it was weird," to which Daria laughs, but TA Joey takes his idea seriously and presses him to clarify. Daria offers that "Maybe it's just getting started up or something," to which Chrissie laughs, but TA Joey again takes the idea seriously and considers a consequence of the idea "so if we did it again…". Finally, Chrissie declares "I think we should do a second trial, to see" and Brandi offers another reason why a second trial would help ("Maybe he wasn't walkin' that steady").

At this point that TA Joey comments on this sensemaking discussion in order to make an explicit point about what it is the group should be doing in tutorial (Figure 9).

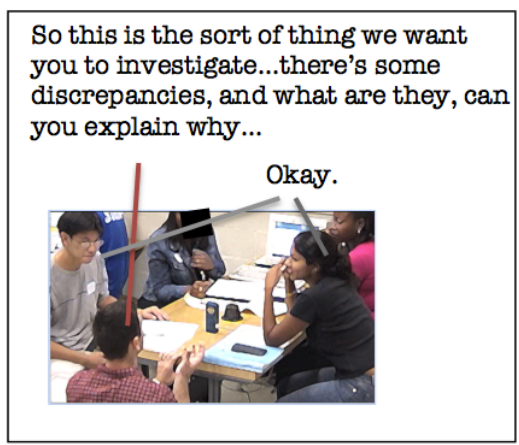

Figure 9—TA Joey uses this sensemaking discussion as an opportunity to repair the Silver group's understanding of what they should be doing in tutorial.

After TA Joey leaves, the Silver group does not go back to their dismissive approach to the tutorial. Instead, they continue to sensemake about the causes of the jumps in the graph. First, they follow Chrissie's suggestion and do another trial:

>             DARIA:  D'you wanna try again?
>              ALAN:  You wanna try it again?
>           CHRISSIE:  Yeah
>             DARIA:  Yeah I just wanna try
>            BRANDI:  *(looking at computer screen)* How did…
>             DARIA:  Hold on,
>           CHRISSIE:  You gotta stand in front of it…ready?
>              ALAN:  Yep *(walks slowly away with book in hand)*
>             DARIA:  *(looking at the new graph)* THERE you gooooo!
>           CHRISSIE:  Ahhhh, okay!
>    BRANDI & DARIA:  *(laughing)*
>             DARIA:  Okay it worked out.

After trying it again and finding a straight line with no jumps in it, the Silver group celebrates with smiles and laughter, saying "Okay it worked out." Even though it seems there is resolution and they can move on, the Silver group continues their sensemaking discussion as they try to resolve the discrepancy between the first and second trial:

>           CHRISSIE:  So maybe you weren't walkin' at a steady pace at one point,
>              ALAN:  Probably, I probably like moved the book or something like that
>             DARIA:  Did you? Yeah maybe
>              ALAN:  Yeah.
>             DARIA:  Wait did you do something different the first time?
>              ALAN:  No.
>             DARIA:  Like, while you were walking back?
>              ALAN:  I was- I prob'ly…I donno either=
>            BRANDI:  Sometimes you do things subconsciously
>              ALAN:  =moved the book down or, you know, yeah.
>           CHRISSIE:  So where do you, you write that where? Oh. B.

Overall the Silver group's discussion provides evidence that the TA's intervention has had a lasting effect on the their understanding of their activity, at least on a short timescale.[3] In this discussion, the Silver group illustrates that they have learned that the "jumps" in the graph are entities they should point out and try to make sense of. This was facilitated when TA Joey overheard a good question, and then incrementally introduced epistemic distancing to encourage the students to offer ideas to explain their graph in terms of physical phenomena. The students themselves used epistemic distancing as they offered ideas half-jokingly, although the TA took them seriously. TA Joey's interaction helped repair the Silver group's epistemological understanding of what it is they are supposed to be doing in tutorial.

### The Bronze group's 1st collaborative sensemaking discussion

Part I demonstrated that the Bronze group's first discussion contained some of the precursors of collaborative sensemaking. For instance, Cathy and Devin each described some of the mechanisms by which talking about your mistakes could help them learn. But Adam's comment, "It's been proven that you learn from your mistakes" seemed to shut down the conversation. The Bronze group's discussions contained little evidence of collaborative sensemaking for the rest of the first tutorial.

---

[3] In fact, the Silver group continue to sensemaking about their motion graphs, so much so that later in the tutorial they sensemake about bumps in their graphs even when the tutorial worksheet tells them to just "smooth out the bumps."

It is not until the beginning of the second tutorial that the group goes into a discussion in which they are collaboratively making sense of a physical phenomenon. In Tutorial 2, Cathy is gone and a new member Britte is present. Whether she missed the first tutorial or was just with a different group, Britte is not yet familiar with the Bronze group's dismissive approach to tutorial. As in the first tutorial, the second tutorial asks the students to make motion graphs and to compare their predictions against the resulting graph. In Tutorial 2, they are making graphs of their *velocity* versus time. The Bronze group starts out the tutorial drawing their predictions of a velocity vs. time graph for someone walking slowly away from the detector. They are focused silently on their worksheets for several minutes, before Brad suggests they get to the experiment:

> BRAD: Should we let it rip?
> BRITTE: Are we um, allowed to discuss now?
> DEVIN: Yes.
> BRITTE: Mmkay...let's see...

Brad suggests they get started with the experiment. Britte, who is new to the group, makes a bid to discuss by using considerable epistemic distancing: "Are we, um allowed to discuss now?" Devin answers in the affirmative, and this prompts them to show each other their graphs and to discuss their predictions:

> BRAD: *(holds his tutorial worksheet up, silently, for others to see)*
> DEVIN: Wait... *(places her worksheet in the middle of the table)*
> BRITTE: *(looks at Devin's worksheet, holds hers up)* I have the opposite of you aheh...Why?
> DEVIN: *(looks at Brad's worksheet)* So, I guess my thinking was the um...velocity's gonna increase *(gestures path of cart down the ramp with hand, down & to the right)* AS it's going down?
> ADAM: But since it's a constant acceleration wouldn't it be a *(gestures a line up and to the right)*
> BRAD: Well, velocity's gonna increase *(gestures straight line up and to the right)* because, it's just FALLing *(repeats gesture up and to the right)*...slower, so things...increase steadily in speed when they fall. And they fall at constant acceleration *(repeats gesture again)*.
> ADAM: Constant acceleration but shouldn't the velocity...curve...
> BRAD: Yeah so velocity is positive...
> ADAM: *(gestures curve with fingers slightly curled)* be a curve as opposed to a straight line *(straightens fingers)*?
> DEVIN: Right right.
> ADAM: 'Cuz the velocity's going to *(traces a curve in the air that flattens out)*
> DEVIN: Level off *(mirrors Adam's gesture)*
> BRITTE: You sure it's not the opposite? Why am I thinking it's the opposite?
> ADAM: But you don't change your velocity though. 'Cuz accelera- 'cuz it's constant acceleration, should have a change in velocity.
> BRAD: *Should have*, or shouldn't.

```
ADAM:  It should.
BRAD:  It's constant acceleration, velocity should- yeah it'd just be a straight
       line.
ADAM:  Oh it's a straight line? (pauses, then nods)
BRAD:  Should we- Should we drop it and try it and see what we got?
ADAM:  (nods again)
```

This discussion contains evidence that the Bronze group is finally able to make space to collaboratively make sense of a phenomenon. They notice inconsistencies in their predicted graphs and seek to resolve them by reasoning about how the physical motions connect with features of the graph. Britte seems to have drawn a graph that represents the physical path of the cart down the ramp, rather than the increasing values of its velocity. Adam and Brad both think the graph should go up, but disagree on whether it should be a straight line or curved. By the second time Brad suggests they try it out, they have a legitimate controversy to settle. If they had tried it out the first time Brad suggested it, they likely would never have noticed their disagreement, let alone discussed it.

There is evidence that the Bronze group's sensemaking here is facilitated in part through the use of epistemic distancing. When Britte challenged Brad's initial move to try it out by suggesting that they discuss their predictions, she did so with considerable epistemic distancing. First, she phrased her request as seeking permission ("Are we, um, allowed to discuss now?"). Her pitch rose significantly by the end of her question, denoting uncertainty (Ward & Hirschberg, 1985). And as she asked her question she pushed her body away from the table, physically distancing herself from the group (Figure 10).

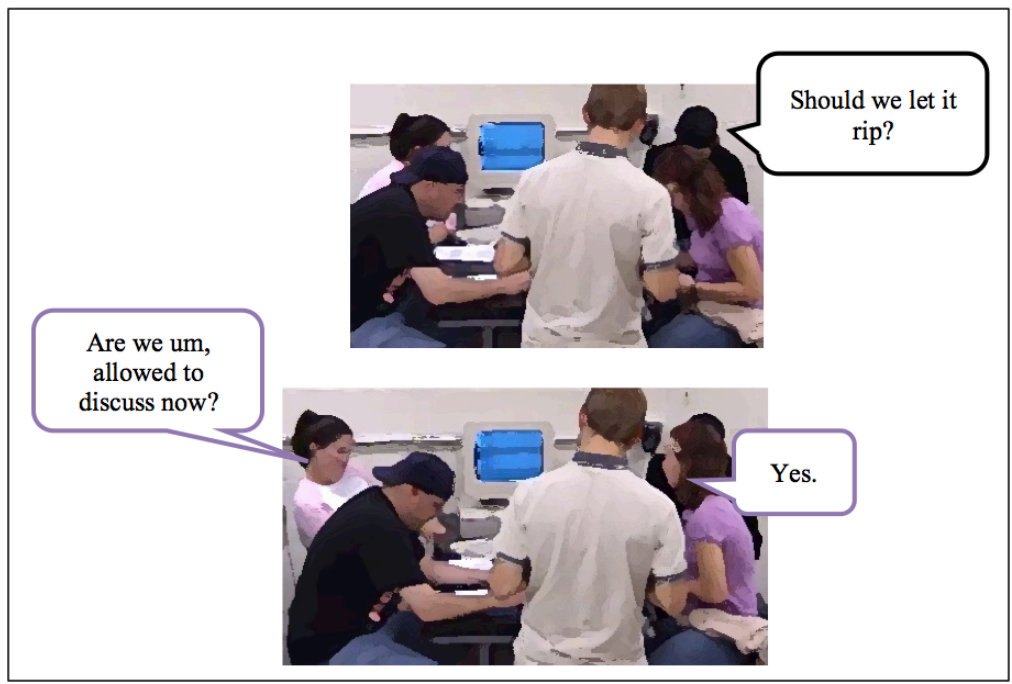

Figure 10—Brad bids to start the experiment, while Britte suggests that they discuss their predictions, with considerable epistemic distancing.

Phrasing her question in this way created an opening for Devin to confirm her request to discuss, after which the group proceeded to discuss their ideas rather than jumping right to the experiment as Brad had suggested. When Britte asked why they have opposite graphs, Devin used epistemic distancing by hedging her idea ("I guess") and explaining what she was thinking ("my thinking was") rather than presenting it as something she *knows*. Adam disagreed with her, but used epistemic distancing by phrasing his disagreement as a question ("but since it's constant acceleration wouldn't it be a") before gesturing a line going up to the right instead of Devin's line going down.

Finally, during the second tutorial of the semester, we have evidence that the group has made a safe space to sensemake, i.e., to put their ideas out there, even if they disagree with others. And once again, the group's establishment of this space to sensemake depended sensitively on the use of epistemic distancing.

Once they have this first sensemaking discussion, the Bronze group continues to have them throughout the rest of the semester, but not as frequently as the two other groups. The Bronze group's discussions also tend to be less mechanistic in nature than the other groups', a contrast exemplified in the comparison of the Gold Group's & Bronze group's loop-the-loop discussions in the Introduction.

### Summary of Part II – Making Space for Sensemaking

Part II explored the dynamics leading to each group's first collaborative scientific sensemaking discussion, finding that in each case epistemic distancing played a critical role. The Gold group made steady progress towards making a safe space for sensemaking, in part by using epistemic distancing in ways that help maintain a safe space to introduce their own ideas and to evaluating them. In contrast, the Silver group seemed to be heading into anti-alignment with the goals of the tutorial, when a TA stepped in at the right time and supported their collaborative sensemaking by incrementally adding epistemic distance to his questions. And finally, the Bronze group did not collaboratively sensemake together until a new student asked a question in with enough epistemic distance ("Are we um, allowed to discuss now?") that the norms of the group shifted towards sharing and evaluating each other's ideas. In all three groups, epistemic distancing played a critical part in the groups' finding a safe space to sensemake.

### Summary

In science, there is an essential tension between the generation of new ideas and the critical evaluation of those ideas (T.S. Kuhn, 1977). This tension is present in active engagement science classrooms that focus on learning through authentic scientific practices (Ford, 2008). This paper demonstrates how three student groups in introductory physics tutorials were able to able to navigate this essential tension in building a safe space to collaboratively make sense of mechanisms. They do so, in part, through the use of epistemic distancing – softening their stances through hedging, joking, quoting, and other shifts of footing. By distancing themselves from the substance of their talk, the students can protect themselves from the affective risks of having their ideas critically evaluated. This in turn can create a safe space where the generation and evaluation of ideas is welcomed. A close discourse analysis revealed how three tutorial groups managed these dynamics within their first few discussions of the semester. A core finding is that epistemic distancing played a key role in each groups' construction of a safe space to collaboratively make sense of mechanisms.

Part I analyzed each group's very first discussion of the semester, finding that all the groups got into the discussion by following the tutorial instructions. However, their different approaches provide evidence of variation across groups in how they understood those instructions. Part of this variation can be explained by differences in how each group managed the epistemological and affective dynamics of their discussion. Through the use of epistemic distancing, the Gold group was able to construct the beginnings of a safe space to sensemake as they shifted from reading what they wrote to saying what they think. The other groups either distanced themselves too much (the Silver group) or too little (the Bronze group) to share much of their thinking in the first discussion.

Part II identified and analyzed each group's first collaborative scientific sensemaking discussion, i.e., the first discussion that showed significant evidence that the students were collaboratively making sense of physical phenomena. There was variability in the timing of when in the tutorials this discussion occurred for each group, and how they got into the discussion. In the case of the Gold group, they progressed steadily in their sensemaking over the first few discussions. For the Silver group and Bronze group, the groups were showing few signs of sensemaking, until an outsider from the group challenged them to engage in the tutorial in a new way. The Silver group started sensemaking together after a nearby TA overheard a good question and used as an opportunity to engage the group in a discussion, which he managed through the use of epistemic distancing. He also used this as an opportunity to repair the Silver group's understanding of what they are supposed to be doing in tutorial. The Bronze group started sensemaking when a new member asked the group, "Are we um, allowed to discuss now?" The request to discuss challenged the group's norms, but was asked with enough epistemic distancing that the group took it up. They started sharing and evaluating their own ideas, even when these conflicted with each other's.

While all of the groups eventually managed to create safe space to sensemake together, it took some groups longer than others to do so. Part of this variability can be explained by differences in the groups' uses of epistemic distancing. The Gold group used epistemic distancing in their first discussion, and made steady progress until by their third discussion they were making sense of physics together. Meanwhile, the Silver group initially distanced themselves too much from the activity to welcome further discussion ("Whatever…Next!"), until TA Joey introduced epistemic distancing into his questions to solicit their ideas. In contrast, the Bronze group used too little epistemic distancing to welcome further discussion ("It's been proven that you learn from your mistakes"), until a new member challenged the group's norms by introducing epistemic distancing into her request to discuss their ideas.

**Implications For Research & Instruction**

These findings build upon research that has explored how students in active engagement science classrooms come to understand the epistemological nature of their activity, i.e., their epistemological framing (Conlin, Gupta, Scherr, & Hammer, 2007; Hammer, Elby, Scherr & Redish, 2005; Redish, 2004; Scherr & Hammer, 2009; Berland & Hammer, 2012). Specifically, it demonstrates how groups come to frame their activity as an opportunity to have collaborative scientific sensemaking discussions, despite the risk of embarrassment that comes with sharing and evaluating each other's ideas.

While research on learning science through inquiry has demonstrated the importance of argumentation and critique (Berland & Hammer, 2012; Kuhn & Pease, 2008; Osborne, 2010), very little research has attended to the affective dynamics of argumentative discussion. The present work highlights that for these tutorial groups their productive sensemaking is not a matter of purely conceptual or epistemological dynamics of the group: There is an undeniable affective component, which can be weighed in the embarrassment that comes from opening up and sharing an idea, especially if that idea is rejected. An idea can be a conceptual in that it pertains to the physical phenomena, as in Deirdre's idea about the normal force ("Would it be going up or would it be going like, like [up at an angle]?"). An idea could also be epistemological, in that it pertains to how they should go about learning together, such as Bree's "Discusssss our answersss" or Britte's "Are we, um, allowed to discuss now?".  Either way, the embarrassment of having one's idea rejected –- or the avoidance of that embarrassment—can shut down collaborative sensemaking.  Here we have found how groups of students navigate these conceptual, epistemological, and affective dynamics all at once, through the use of epistemic distancing. By hedging, joking, or other means of softening their stance, speakers can create a buffer between the person and the idea, so that the idea can be evaluated rather than the person.

**Implications For Instruction**
The finding that students and instructors alike spontaneously use epistemic distancing to successfully set up a safe space has direct instructional implications.  Epistemic distancing moves could prove useful to instructors and curriculum designers looking to support students' collaborative scientific sensemaking. The data presented above illustrate how a subtle shift in how an instructor words a question, from "What happened there?" to "What do you *think* happened there? Any idea?" can have immediate consequences on students' willingness to share their ideas. Deirdre's suggestion for the Gold group to start with reading what they wrote assuaged the Gold Group's discomfort with discussing their ideas. This move could easily be adopted by an instructor or even a curriculum developer interested in supporting students' sensemaking [4].

    Before rushing from observation to prescription, however, it is crucial to emphasize two points. First, more epistemic distancing is not necessarily better. The Silver group started off using too much epistemic distancing to have productive discussions, while the Bronze group used too little. Second, moves like TA Joey's and Deirdre's were constructed on the spot *in response to* the ongoing activity.  For some groups, asking "What happened there" may be enough to get them discussing their ideas about mysterious jumps in a graph, but TA Joey evidently noticed in the moment that the Silver group needed more support in sharing their ideas. Such in-the-moment instructional moves require attending and responding to the students' affect, especially their comfort with sharing ideas.  It could be that later on in the semester, after the Silver group had already established a safe space to sensemake, less epistemic distancing would be needed to support productive discussion. These longer-term dynamics warrant further study.

**Discussion – Degrees of Belief in Science**
    To close, it is important to address one essential tension that has gone unmentioned: Is the use of epistemic distancing consistent with doing "good science"?  In science, a hypothesis must take risks (Godfrey-Smith, 2003, p. 58). It must "stick its neck out" so we can put it to a stringent test, and only accept it once it has "proved its mettle" (Popper, 2005, p. 32).  If science favors bold claims, isn't epistemic distancing to be avoided in science, and therefore in science classrooms?  Such a view becomes untenable when taken to the extreme. The history of science teaches us that ideas that once proved their mettle can later be rejected, and that once-rejected

---

[4] *The Tutorials in Physics Sensemaking* are open source, so instructors may adjust them to meet their particular needs, say, by adding epistemic distancing into the worksheet questions.

theories can make a comeback, as was the case of the corpuscular theory of light (Lakatos, 1980).  For this reason, it is useful for scientists to hold more nuanced stances towards ideas than mere acceptance or rejection, such as pursuing an idea without necessarily believing in it (Laudan, 1981; Whitt, 1990).

Failure to appropriately manage epistemic distance can pose risks to scientists' careers. When physicists reported the detection of faster-than-light neutrinos at CERN in 2011, they did not boldly claim that they had overthrown the theory of relativity. Instead they noticed the discrepancy with relativity and asked for other teams to attempt a replication (Cho, 2011). Through this process, the cause was found: a mundane case of faulty wiring. Had these scientists gone for the "bold" claim, their careers would now be over, but they understood that the boldness of hypotheses in science should be held in proportion to the strength of supporting evidence and their fit with established theory. The risk-taking environment faced by students in collaborative scientific sensemaking discussions is a microcosm of the risk-taking environment of doing cutting-edge science. In both cases, doing good science involves taking more nuanced epistemic stances towards ideas than mere acceptance or rejection. Doing good science involves the management of epistemic distance.

## Acknowledgments
This work was funded by the National Science Foundation (Grant #0440113).  The authors would like to David Hammer for guidance in this work, Ayush Gupta for assistance with analysis and helpful discussions, as well as Megan Luce, Eric Kuo, and Dan Schwartz for helpful comments and suggestions on earlier drafts.

Making Space to Sensemake - PREPRINT	26Conlin, L.D. (2012). *Building shared understandings in introductory physics tutorials through risk, repair, conflict, and comedy* (Doctoral dissertation.) Available from ProQuest Dissertations and Theses database. (UMI No. 3517523)

Conlin, L.D., Gupta, A., Scherr, R.E., & Hammer, D. (2007). The dynamics of students' behaviors and reasoning during collaborative physics tutorial sessions. In L. Hsu, C. Henderson, & L. McCullough (Eds.), Proceedings of the Physics Education Research Conference (Vol. 24, pp. 66-72). Greensboro, NC: American Institute of Physics.

Duschl, R. (2008). Science education in three-part harmony: Balancing conceptual, epistemic, and social learning goals. *Review of Research in Education, 32*(1), 268-291.

Engle, R.A., & Conant, F.R. (2002). Guiding principles for fostering productive disciplinary engagement: Explaining an emergent argument in a community of learners classroom. *Cognition and Instruction, 20*(4), 399-483.

Ford, M.J. (2008). Disciplinary authority and accountability in scientific practice and learning. *Science Education, 92*(3), 404-423.

Godfrey-Smith, P. (2003). *Theory and reality: An introduction to the philosophy of science*. Chicago: The University of Chicago Press.

Goffman, E. (1955). On face-work: An analysis of ritual elements in social interaction. *Psychiatry, 18*(3), 213-231.

Goffman, E. (1956). Embarrassment and social organization. *The American Journal of Sociology, 62*(3), 264-271.

Goffman, E. (1974). *Frame analysis: An essay on the organization of experience*. Cambridge, MA, US: Harvard University Press.

Goffman, E. (1979). Footing. *Semiotica*, *25*(1-2), 1-30.

Goodwin, C. (2007a). Interactive footing. In E. Holt & R. Clift (Eds.) *Reporting Talk: reported Speech in Interaction* (Vol 24, pp. 16-46). Cambridge, UK: Cambridge University Press.

Goodwin, C. (2007b). Participation, stance and affect in the organization of activities. *Discourse & Society*, *18*(1), 53-73.

Heisterkamp, B. L. (2006). Taking the footing of a neutral mediator. *Conflict Resolution Quarterly, 23*(3), 301–315.

Hirst, G., McRoy, S., Heeman, P., Edmonds, P., & Horton, D. (1994). Repairing conversational misunderstandings and non-understandings. *Speech Communication, 15*(3-4), 213–229.

Holmes, J. (1984). Modifying illocutionary force. *Journal of Pragmatics, 8*(3), 345–365.

Holmes, J. (1990). Hedges and boosters in women's and men's speech. *Language & Communication, 10*(3), 185–205.

Hoyle, S. M. (1998). Register and footing in role play. In S. M. Hoyle & C. T. Adger (Eds.), *Kids talk: Strategic language use in later childhood* (pp. 47–67). New York: Oxford University Press.

Jacobs, S. (2002). Maintaining neutrality in dispute mediation: managing disagreement while managing not to disagree. *Journal of pragmatics, 34*(10-11), 1403–1426.

Making Space to Sensemake - PREPRINT                                                                 28Sacks, H., Schegloff, E.A., & Jefferson, G. (1974). A simplest systematics for the organization of turn-taking for conversation. *Language, 50*(4), 696-735.

Scherr, R. E., & Hammer, D. (2009). Student behavior and epistemological framing: Examples from collaborative active-learning activities in physics. *Cognition and Instruction, 27*(2), 147–174.

Schiffrin, D., Tannen, D., & Hamilton, H.E. (2001). *The handbook of discourse analysis*. Malden, MA: Blackwell Publishers.

Shaffer, P. S., & McDermott, L. C. (1992). Research as a guide for curriculum development: An example from introductory electricity. Part II: Design of instructional strategies. *American journal of physics*, *60*(11), 1003-1013.

Sharma, B. (2011). *Conceding in disagreements during small group interactions in academic writing class*. (Master's Thesis, University of Hawai'i at Manoa). Retrieved from http://scholarspace.manoa.hawaii.edu/handle/10125/20166

Tudge, J. (1989). When collaboration leads to regression: Some negative consequences of socio-cognitive conflict. *European Journal of Social Psychology, 19*(2), 123–138.

Ward, G., & Hirschberg, J. (1985). Implicating uncertainty: the pragmatics of fall-rise intonation. *Language, 61*(4), 747–776.

Whitt, L. A. (1990). Theory Pursuit: Between Discovery and Acceptance. *PSA: Proceedings of the Biennial Meeting of the Philosophy of Science Association*, 1990, 467-483.

Making Space to Sensemake - PREPRINT                                                                29## Appendix A – Transcription Conventions

Transcripts follow a variant of the Jefferson transcription system (Sacks, Schegloff, & Jefferson, 1974, pp. 731-733).

| Sign | Description | Example |
|---|---|---|
| . , ? | Punctuation indicates pitch variation at the end of utterances, not grammar of sentences. | A: I dunno. Maybe, this was weird? <br> D: Hehehe <br> TA: Maybe it was weird. |
| **Boldface** | Indicates emphasis signaled by changes in pitch. | "So just helps you understand the way **you** think of it" |
| CAPITALS | Indicate increased volume. | "THERE you go!!" |
| - | A <u>dash</u> denotes a sudden cut-off of speech. | 'Cuz accelera- 'cuz it's constant acceleration |
| … | <u>Ellipses</u> denote a significant pause in speech. | "I guess we should…what did we have to do?" |
| ssss | <u>Repeated letters</u> denote elongated pronunciation. | Discusss our answerssss |
| (*actions*) | • *Italics* in parentheses indicate actions, including gestures, which accompany the speech. | (*points to worksheet*) |
| Contiguous= =talk | An <u>equals sign</u> is used to indicate "latching"; there is no interval between the end of a prior unit and the start of a next piece of talk. | D: the way **you** thought about it= <br> C: I put…if you <br> D: =versus the correct way |